\documentclass[traditabstract]{aa}
\usepackage{graphicx}
\usepackage{epsfig}
\usepackage{color}
\usepackage{txfonts}
\usepackage{aalongtable}
\usepackage[authoryear]{natbib}
\usepackage{bm}
\bibpunct{(}{)}{;}{a}{}{,}

\begin{document}

\title{A VIMOS spectroscopy of photometric variables and straggler candidates 
       in $\omega$~Centauri 
       \thanks{Based on photometric data collected at Las Campanas Observatory, and spectroscopic 
       data collected with the Very Large Telescope at European Southern Observatory (ESO programme 
       384.D-0736).}}

\titlerunning{VIMOS spectroscopy of $\omega$~Centauri}

\subtitle{}

\author{M. Rozyczka\inst{1}
        \and
        J. Kaluzny\inst{1}
        \and
        P. Pietrukowicz\inst{1,2}
        \and
        W. Pych\inst{1}
        \and
        M. Catelan\inst{2}
        \and
        C. Contreras\inst{2}
        }

\authorrunning{Rozyczka et al.}


\institute{Nicolaus Copernicus Astronomical Center,
ul. Bartycka 18, 00-716 Warszawa, Poland
        \and
Pontificia Universidad Cat\'olica de Chile,
Departamento de Astronom\'ia y Astrof\'isica,
Av. Vicu\~na MacKenna 4860, 
782-0436 Macul, Santiago, Chile
}

\date{Received ...; accepted ...}

\abstract
{
We report a spectroscopic study of 19 photometric variables and 55 blue, yellow and red 
straggler candidates in the field of $\omega$ Centauri. 
We confirm the cluster membership of 18 variables and 
54 straggler candidates. Velocity variations are detected in 22 objects, and another
17 objects are classified as suspected of being velocity-variable. Velocity data of 
11 photometric variables phase with their photometric periods, however none of these
objects has a mass function indicating the presence of a massive degenerate component. 
Based on both photometric and spectroscopic data we find that the fraction of binaries 
among blue stragglers may be as high as 69 per cent.
}

\maketitle

\section {Introduction}
\label{sect:intro}

Globular clusters harbor a massive population of close binary systems with degenerate 
components (hereafter referred to as degenerate binaries), discovered primarily due to 
their X-ray emission \citep[e.g.][and references therein]{poo10}. Unfortunately, detailed 
studies of these interesting objects are hampered by the fact that their optical counterparts 
are often weak and hard to identify in the crowded environment \citep[e.g.]{ver08}. On the 
other hand, many field X-ray binaries exhibit long periods of quiescence. The nature of a
quiescent binary is betrayed by an low-amplitude optical modulation induced primarily by the 
ellipsoidal effect from the nondegenerate component, while a spectroscopic observer would 
see it as as a single-line system with large orbital velocity ($K >150$~km~s$^{-1}$). Perhaps 
the best examples of such objects are X-ray novae which almost certainly harbor black holes. 
They would be distinguished by a mass function $f_m = (m\sin i)^3/(m_{bin})^2 >2M_{\odot}$, 
where $m$ and $m_{bin}$ stand, respectively, for the mass of the degenerate component and the 
total mass of the binary \citep{rem06}.

The first spectroscopic search for quiescent degenerate binaries in globular clusters was 
conducted by \citet{roz10} based on a sample of short-period, low-amplitude optical variables 
with nearly sinusoidal light curves and periods shorter than $\sim$1.3~days. The sample 
included 4 objects from NGC 6397 and 7 objects from $\omega$ Centauri (NGC 5139). The most 
interesting findings were turnoff binaries V17 and V36 in 
NGC 6397, whose invisible primary components had masses larger than 1 $M_\odot$ (in the 
case of V36 even larger than 2 $M_\odot$, albeit with a large uncertainty). In the present 
paper we continue the search for degenerate binaries in $\omega$~Cen based on a sample of 
20 optical variables with periods obtained by \citet{kal04}, which were classified by 
\citet{bel09} as proper-motion members of that cluster. 

Another class of cluster members certainly worth of spectroscopic investigation 
consists of objects populating ``nonstandard'' areas of the color-magnitude diagram (CMD): 
blue stragglers, yellow stragglers, and sub-subgiants \citep[also called red stragglers; see 
e.g.][]{pla11}.

Blue stragglers were originally defined 
as stars located at the extension of the main sequence above the turnoff point. Being both 
brighter and hotter (``bluer'') than turnoff stars, they made an impression of evolving at 
a slower rate than normal cluster members. Although they have been 
known for almost 50 years \citep{san53}, the mechanism of their formation remains 
controversial. The various possibilities include mass transfer between the components 
of a binary, merger of a binary or direct collision of stars. It is also conceivable that 
all these mechanisms are at work, perhaps with various relative eficiency in
various clusters \citep{fer06,fer106,dal08}.  

The results of \citet{roz10} point to an extensive mass exchange rather than stellar
collisions, as in all cases they report the brighter component of the blue-straggler 
binary is the more massive 
primary star which must have acquired significant amounts of hydrogen-rich material from 
its companion. Thus, if their sample is representative, then many blue stragglers 
must be Algol-like systems in which the original mass ratio has been reversed, causing 
the mass transfer to effectively stop. One of the best studied blue stragglers, star V228 
in 47 Tuc is indeed a classical Algol \citep{kal07b}; however, in other cases the similarity 
to Algols may be purely morphological. A good example is the also well-studied star V209 
in $\omega$ Cen, which most probably underwent two mass transfer episodes. Its present 
primary seems to be ``reborn'' from a former white dwarf that accreted a new envelope 
through mass transfer from its companion, while the present secondary lost most 
of its envelope during the ascent along the subgiant branch, failed to ignite helium, 
and is now powered by a hydrogen-burning shell \citep{kal07a}. 

Yellow stragglers reside in the area between the subgiant branch and the horizontal 
branch.
Most probably, they are evolutionary advanced blue stragglers \citep[e.g.][]{xin11}, and 
distinguishing between these two groups of objects seems to be a matter of taste rather 
than physical necessity. Red stragglers are found to the right of the main sequence,
below the subgiant branch. They are even less explored than the blue ones, and the 
conundrums they pose are even more mysterious \citep{pla11}. 

Both the degenerate binaries and all kinds of stragglers are thought to be products of 
the evolution of binary systems in a dense stellar environment, and their peculiarities 
are most probably promoted or even induced by interactions between cluster members 
\citep{fer06,pla11}. As such, they provide a link between classical stellar evolution and 
dynamical 
evolution of the cluster, being a valuable observational template against which 
the dynamical models of stellar aggregates can be tested. The straggler-related goal of 
the present survey is to verify the membership of straggler candidates selected from 
the proper-motion catalogue of $\omega$ Cen by \citet{bel09}, and to establish how frequent 
binary systems are in the straggler population. 

\section {Observations and data reduction}

We monitored selected targets in $\omega$ Cen with the help of VIMOS -- a multi-purpose 
instrument mounted in the Nasmyth B focus of the ESO VLT-Unit 3 telescope. For the present
survey it was working as a multi-object spectrograph. To adapt it to the observations of blue 
stragglers, we selected the HR blue mode with wavelength range 4100 -- 6300 
\AA, resolution 2050 -- 2550 (150 -- 120 km~s$^{-1}$) and dispersion 0.5~\AA/pixel.  

A VIMOS spectroscopy run consists of pre-imaging and spectroscopic follow-up. The 
pre-imaging frames of $\omega$ Cen were obtained on 2010.16.01 to serve as a basis for 
the preparation of masks with slits centered on objects chosen for the survey. The 
spectroscopic monitoring was performed during ten nights in February and March 2010. On every 
night one $\sim$30 min observation was made, consisting of acquisition-imaging, two 
spectroscopic integrations of 580 sec. each, up to three flat-field exposures, and 
helium-neon lamp exposure for wavelength calibration. The VIMOS field of view, which 
is composed of four 7$\times$8 arcmin quadrants served by independent CCDs and separated 
by about 2 arcmin gaps, was centered on ($\alpha, \delta)_{2000}$ = (13$^{\rm h}$26
$^{\rm m}$46.3$^{\rm s}$, -47$^\circ$27\arcmin51.8\arcsec). A log of the observations 
is presented in Table \ref{tab:log}, in which date and airmass are given for the start 
of the exposures, and the seeing is averaged over each observation. 

\begin{table}[h!]
  \caption{Log of observations.  
           \label{tab:log}}
   \begin{tabular}{rccc}
    \hline
     ID&UT date&Airmass&Seeing (arcsec)\\
     \hline 
     \hline 
      1&2010 02 11.32 &1.11 &2.0\\
      2&2010 02 12.32 &1.11 &1.5\\
      3&2010 02 13.31 &1.12 &1.0\\
      4&2010 02 14.30 &1.14 &1.0\\
      5&2010 02 21.26 &1.19 &1.0\\
      6&2010 02 22.30 &1.11 &0.7\\
      7&2010 02 23.25 &1.19 &0.7\\
      8&2010 03 09.38 &1.21 &1.0\\
      9&2010 03 10.30 &1.09 &1.3\\
     10&2010 03 12.24 &1.11 &0.7\\
     \hline 
   \end{tabular}
\end{table}

The target objects were selected from variable star catalog of \citet{kal04}, 
henceforth identified by a number preceded with V or NV, and proper-motion 
catalog of \citet{bel09}, henceforth identified by a number preceded with B. 
We adopted the following selection criteria:
\begin{itemize}
\item $V < 19$ mag (assuming 1\arcsec\ slit, 1\arcsec\ seeing and reasonably 
long exposure of $\sim$1000 s, this is the magnitude at which the $S/N$ ratio 
expected for the latest spectral types in our sample amounts to $\sim$20);
\item Membership probability $mp$ in \citet{bel09} should be at least 90\%; 
\item The image of the target should not blend with images of other objects on 
pre-imaging frames;
\item The distribution of targets on pre-imaging frames should maximize the number 
of slits. 
\end{itemize}

Photometrically detected pulsating stars, whose intrinsic variations of radial 
velocity could mask orbital effects, were excluded from the sample.
The selection procedure was rather tedious, and trying to maximize the number of 
photometric variables in the sample we were forced to relax the second criterion. 
After a few trials we decided to focus on 81 objects: 61 straggler candidates and 
20 photometric variables, among which NV332, NV334 and NV361 had membership 
probabilities equal to 84\%, 88\% and 64\%, respectively. We deliberately included 
V209 -- a system thoroughly investigated by \citet{kal07a} -- with the intention
to use it as an indicator of the quality and reliability of velocity measurements.
In the sample there are a few blue objects which, strictly speaking, do not 
conform to the classical definition of blue stragglers, because they are located 
not on the main-sequence extension, but to the left of it. We included them in order
to maximize the total number of slits.

VIMOS spectra can be calibrated automatically by the ESO-VIMOS pipeline,
which, however, cannot be 100\% trusted. This is because the spectra of some slits extend below 
$\lambda=5000$ \AA, where calibration lamp lines are scarce \citep{eso11}, and some of them
are very weak. While 
\citet{giu10} wrote an interactive procedure within the pipeline which allowed for a better 
control of each calibration step, we decided to reduce the data manually, using standard 
IRAF\footnote {IRAF is distributed by the National Optical Astronomy Observatories,
 which are operated by the AURA, Inc., under cooperative agreement  with the NSF.} 
procedures. We started from sci\_raw frames and proceeded through object identification, 
slit extraction, flatfielding, science aperture extraction, lamp aperture extraction, 
wavelength calibration and normalization of the spectra. Unfortunately, all flat-field 
images in quadrants 3 and 4 were contaminated 
by internal reflections. We were forced to substitute the affected image-sections with 
smooth fits, which, of course, degraded the quality of the corresponding sections of 
the spectra.  

\begin{figure}[t!]
\centering
\includegraphics[width=0.49\textwidth,bb= 35 405 565 692,clip]{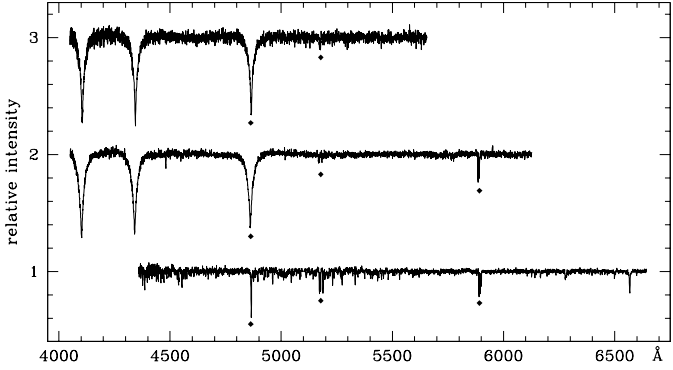}
\caption {Three examples of our spectra. The nominal spectral range of VIMOS in the setup used 
          for the present survey begins at 4100 \AA, but the recorded spectral range depends on 
          the location of the slit on the mask and may extend shortward of that limit. From top 
          to bottom, shown are: B289620 (shifted upward by 2), B319396 (aka V192; shifted upward 
          by 1) and B155556 (aka V216). Diamonds mark lines used for velocity measurements: 
          H$_\beta$ at 4861 \AA, magnesium triplet at 5167-5184~\AA\ and sodium doublet at 
          5890-5896 \AA.} 
\label{fig:example_spectra}
\end{figure}

Seven slits produced no useful data. In five cases the target aperture could not be 
reliably extracted or the spectrum was too noisy, and the catalogued $(B-V)$ values of 
straggler candidates B283374 and B309015 turned out to be entirely inconsistent with 
their spectra, suggesting that those objects were either misidentified or very tightly 
blended. As a result, our sample shrunk to 19 photometric variables and 55 straggler 
candidates listed in Table \ref{tab:object_list}. For each of them we obtained ten 
reduced spectra. Three examples of the spectra are shown in Fig. \ref{fig:example_spectra}.
As detailed below, the information about our objects was mainly extracted from H$_\beta$.
The average S/N at H$_\beta$ was for all objects larger than 20. Out of the total of 740 
spectra only a few had S/N $<20$.   

\begin{figure}[t!]
\centering
\includegraphics[width=0.49\textwidth,bb= 35 405 565 692,clip]{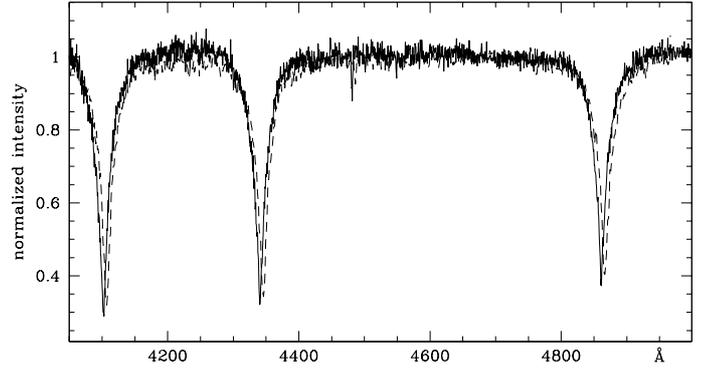}
\caption {A test of the quality of our spectra. Shown is the segment of the spectrum 
          of B319396 (aka V192) common to the present survey (solid line) and that of
          \citet{vl07} (broken line, shifted longwards by 5 \AA\ for clarity). Visible
          are H$_\delta$, H$_\gamma$ and H$_\beta$.} 
\label{fig:spectra_comparison}
\end{figure}

A broad spectroscopic survey of $\omega$ Cen down to $V=16.5$ mag conducted by \citet{vl07}
at a resolution $R\sim$2000 enabled us to verify the quality 
of the reduced spectra by a direct comparison. Among 15 of our variables and straggler 
candidates with $V<16.5$ we found four in common with their sample. In all four cases the 
agreement was good -- an example is shown in Fig. \ref{fig:spectra_comparison}. Note 
that the nominal spectral range of VIMOS in the setup used for our observations starts 
at 4100~\AA\ while the spectra of \citet{vl07} extend from 3840 \AA\ to 4940~\AA, so that the
range covered by both surveys is only 840 \AA\ long.   

\section {Analysis and results}
\label{sect:results}

The actual spectral range recorded by VIMOS depends on the location of the slit on 
the mask. In our data common to al slits was the region between 4700 and 5600 \AA\ 
which in many spectra contained practically no lines except H$_\beta$ and magnesium triplet 
at 5167.32, 5172.68 and 5183.60 \AA\ (in the hottest objects even the latter was practically 
undetectable). The sodium doublet at 5889.95 and 5895.92 \AA\ fell beyond the red end of 11 
spectra, and H$_\alpha$ was visible in just 13 spectra. As one can see, the material for 
radial velocity measurements was highly diverse. Moreover, the accuracy of wavelength 
calibration rapidly deteriorated below 4500 \AA, where only two lamp lines (He 4471.48 \AA\ 
and He 4026.19 \AA) were available. Keeping this in mind and aiming to make our survey as 
uniform as possible, we decided to base the measurements on H$_\beta$ fitting, using other 
means for checkup only.

The velocity was measured with the help of the IRAF task SPLOT by fitting 
Voigt profiles to H${_\beta}$. Wherever possible, it was also measured the same way from 
Mg and Na lines.
For each line the object's velocity was calculated as a mean 
\begin{equation}
 \bar{v} =\frac{1}{N}\sum_{i=1}^N v_i,
 \label{eq:vmean_def}
\end{equation}
where $N\le10$ is the number of fitted spectra, and the corresponding rms deviation 
\begin{equation}
 \sigma=\sqrt{\frac{1}{N}\sum_{i=1}^N(\bar{v}-v_i)^2}
 \label{eq:vsigma_def}
\end{equation}
was found. In the following, velocities and rms deviations are subscripted with the 
symbol of the element from which they are obtained; e.g. $\bar{v}_{\rm H}$ or 
$\sigma_{\rm Na}$. The results of sodium-based measurements are further differentiated 
by indices referring to the origin of the line; e.g. $\bar{v}_{\rm Na,i}$ and 
$\bar{v}_{\rm Na,s}$ are velocities obtained, respectively, from interstellar and stellar 
component of the line. 

We also attempted to measure the velocities with the help of the IRAF task FXCOR. 
The measuring procedure consisted of the following steps: 
\begin{itemize}
\item 
 Based on the recent calibration of \citet{cas10}, the temperature $T$ of each object was 
 estimated from the dereddened $(B-V)$ index assuming $E(B-V) = 0.08$ \citep{mcd09}. 
\item  
 $T$ was rounded to the nearest multiple of 250 K, and a corresponding template from 
 the library compiled by \citet{mun05} was assigned to the object. Both $\log g$ 
 and [Fe/H] were the same for all templates and equal to 4.0 and -1.5, respectively.
\item 
 Each observed spectrum of the object was cross-correlated with the object's template. 
\end{itemize}
The second step may seem oversimplified because of the large chemical composition spread 
in $\omega$ Cen \citep[e.g.][]{joh10} and obvious spread of gravitational acceleration 
across our sample, in which $V$ varies 
by as much as 1.5 mag at constant $B-V$. Also, the empirical formula of \citet{cas10} 
loses validity for $(B-V)<0.19$, i.e. for the eight hottest among our objects, and it may  
not be applicable to the 14 coolest objects lying far to the right of the main sequence. 
However, we checked on several objects that reasonable changes in [Fe/H], $\log g$ 
or even $T$ (i.e. by $\pm0.5$ dex, $\pm0.5$ dex, and $\pm250$ K, respectively) did not 
modify the results of velocity measurements in any significant way. 

The spectral region selected for cross-correlation was located between H$\beta$ 
and sodium doublet (including H$\beta$ or more hydrogen lines resulted in extremely broad 
correlation peaks for hotter objects, whereas including sodium doublet introduced a signal 
from strong interstellar lines). It always contained 
the magnesium triplet, and its extent was chosen so as to maximize the correlation peak. 
Because in many hotter spectra the magnesium lines were not very much stronger than the noise, 
the resultant velocity measurement was considered satisfactory only if 
the maximum value of the correlation function $f^c_m$ exceeded 0.2 (for the synthetic 
template cross-correlated with itself $f^c_m$ was equal to 0.8). 

\begin{figure}[t!]
\centering
\includegraphics[width=0.49\textwidth,bb= 40 160 565 697,clip]{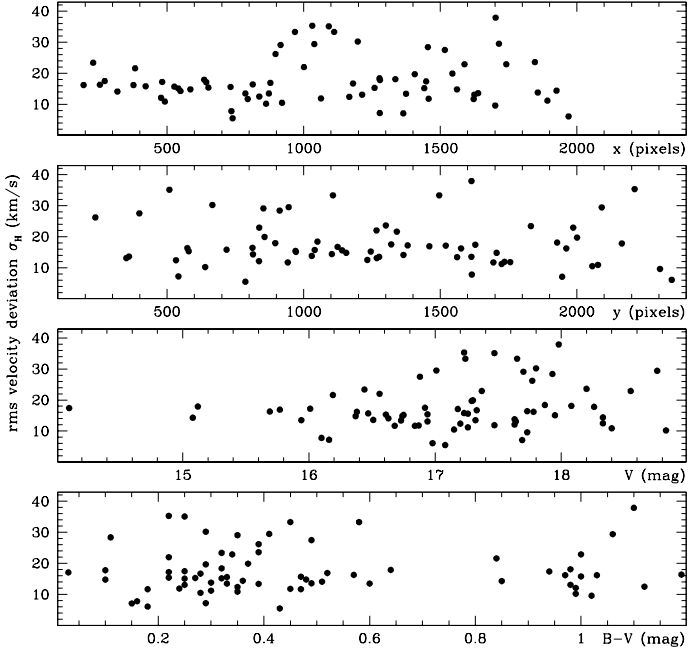}
\caption {Results of velocity measurements based on H$_\beta$ fitting. The rms velocity deviation 
          from the mean velocity of each object is plotted as a function of (x,y) coordinates
          on the VIMOS mask, object's $V$-magnitude and object's color. Three objects with 
          $\sigma_H > 40$ km s$^{-1}$ are not shown in order to better visualize the lower rms-range.}
\label{fig:univ_sigma}
\end{figure}

\subsection{Discussion of errors}
\label{sect:errors}

H$_\beta$ was fitted in all spectra of all objects, yielding 74 mean velocities 
$\bar{v}_{\rm H}$ and 74 corresponding rms deviations $\sigma_{\rm H}$.
The latter originate from inaccurate fitting and/or calibration, and, in the case of binary 
systems, from the orbital motion. We checked that $\sigma_{\rm H}$ does not correlate with 
object's magnitude or color, nor with the location of the corresponding slit on the VIMOS mask
(Fig. \ref{fig:univ_sigma}). 
This allows us to assume that fitting and calibration errors are purely random with 
the same Gaussian distribution for all objects: 
\begin{equation}
 p(x) = \frac{1}{\sqrt{2\pi s^2}}\exp\left(-\frac{x^2}{2s^2}\right),
 \label{eq:std_gauss}
\end{equation}
where $s$ is the (yet unknown) standard error of a single H$_\beta$-measurement. Then our 
observable $\sigma_{\rm H}$ is obtained by randomly drawing ten values from the distribution 
(\ref{eq:std_gauss}), and calculating the root mean square:
\begin{equation}
 \sigma_{\rm H}=\sqrt{\frac{1}{10}\sum_{i=1}^{10}x_i}.
 \label{eq:rms}
\end{equation}
The expected distribution of the variable $y\equiv\sigma_{\rm H}^2$ is chi-square with ten degrees 
of freedom:
\begin{equation}
 p(y)=\frac{26.042}{s^{10}}\exp\left(-\frac{5y}{s^2}\right)y^4.
 \label{eq:chisq}
\end{equation}

\begin{figure}[t!]
\centering
\includegraphics[width=0.49\textwidth,bb= 35 385 565 692,clip]{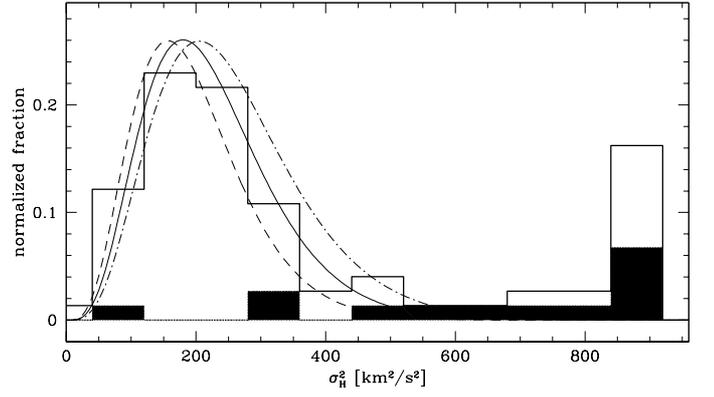}
\caption {Histogram of squared rms velocity deviations obtained from H$_\beta$ fitting. 
          Shaded are photometric variables with periods shorter than 3~d. 
          The expected distributions resulting from equation (\ref{eq:chisq}) for $s$ = 
          14, 15 and 16 km s$^{-1}$, where $s$ is the standard error of a single measurement,
          are shown with dashed, solid, and dot-dashed line, respectively. The rightmost bar
          contains all objects with $\sigma_{\rm H}>29$ km s$^{-1}$.}
\label{fig:chi_sigHb}
\end{figure}

\begin{figure}[t!]
\centering
\includegraphics[width=0.49\textwidth,bb= 35 385 565 692,clip]{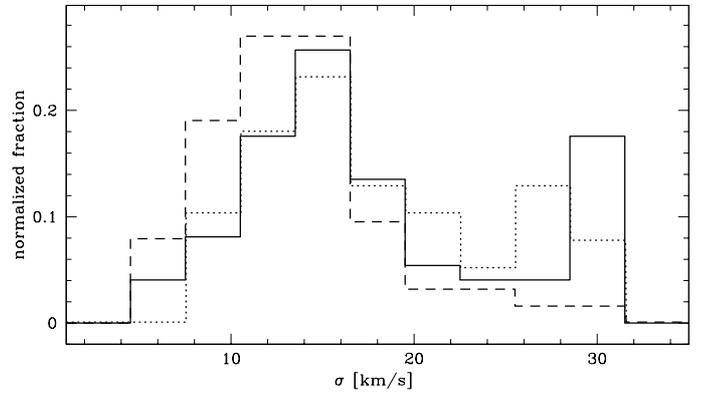}
\caption {Histograms of rms velocity deviations $\sigma_{\rm H}$ (solid), $\sigma_{\rm Na,i}$
          (dashed) and $\sigma_{\rm Mg}$ (dotted). The rightmost bar of each histogram 
          contains all objects with $\sigma>\sim$29 km~s$^{-1}$. $\sigma_{\rm Na,i}$ 
          measurements are based on the interstellar D$_2$ line.}
\label{fig:sigHb_sigNa}
\end{figure}

In Fig. \ref{fig:chi_sigHb} the histogram of $y$ obtained from our measurements is compared to 
distributions (\ref{eq:chisq}) with $s$ =  14, 15 and 16 km~s$^{-1}$. One can see that for 
$\sigma_{\rm H}^2<\sim$400 km$^2$ s$^{-2}$ the observed distribution is quite similar to the 
expected ones, and the hig-velocity tail, clearly visible in the histogram but entirely absent 
from the expected distributions, is largely 
populated by short-period binaries. These two observations indicate that the assumptions
about random nature of H$_\beta$-measurement errors and universality of their distribution are 
entirely reasonable. Among the three fits to the histogram the best one is that with $s=15$ 
km~s$^{-1}$. Thus, we find that the standard error 
$\Delta v_{\rm H}$ of a single velocity measurement based on H$_\beta$ fitting is equal to 
15$\pm$0.5 km~s$^{-1}$. A similar accuracy of velocity measurements from VIMOS spectra 
(10 -- 20 km~s$^{-1}$) was reported by \citet{giu10}. Quantitatively the same, but intuitive 
rather than rigorous estimate of $\Delta v_{\rm H}$ follows from Fig. \ref{fig:sigHb_sigNa}, 
in which the histogram of $\sigma_{\rm H}$ peaks at $\sim$15~km~s$^{-1}$. 

The stellar Na doublet was usually blended with the interstellar one in such a way that the 
velocity could be reliably measured only from interstellar D$_2$ and stellar D$_1$ line
\citep[see also][]{vl07}. The interstellar line was measured in all 63 objects 
whose spectra included the Na doublet, yielding the histogram of $\sigma_{\rm Na,i}$ shown in 
Fig. \ref{fig:sigHb_sigNa}. By analogy with the histogram of $\sigma_{\rm H}$, its central value 
of $\sim$13~km~s$^{-1}$ is a good estimate of the standard error of a single velocity measurement 
based on that line. It is slightly smaller than $\Delta v_{\rm H}$ because 
hydrogen lines are very broad in most objects, which results in increased fitting errors. Also, 
the wavelength calibration is on the average sligthly less accurate at H$_\beta$ than at the 
Na doublet because there are fewer lamp lines in the bluer part of the spectrum. 

\begin{figure}[t]
\centering
\includegraphics[width=0.49\textwidth,bb= 58 322 323 695,clip]{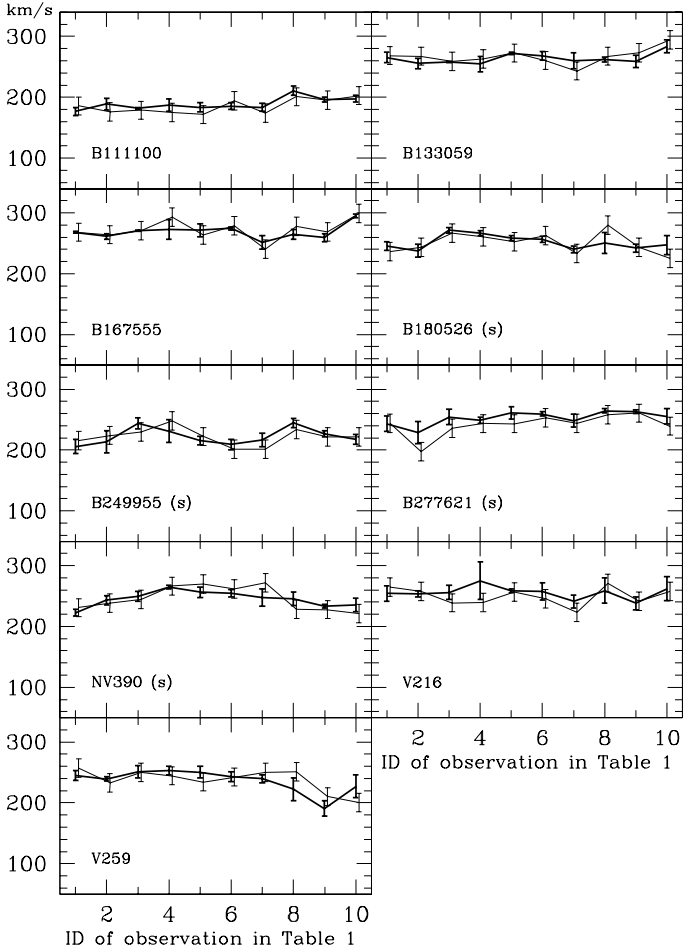}
\caption {Velocities obtained from H$_\beta$ fitting (thin lines) compared to averages $v_4$ 
          defined by equation (\ref{eq:v4}). Heavy errorbars are rms deviations from $v_4$. 
          Thin errorbars (shifted by 0.1 to the right for clarity) mark the standard error 
          of a single H$_\beta$ measurement estimated in Sect \ref{sect:errors} (15~km~s$^{-1}$).
          Objects suspected of being velocity-variable (see Sect. \ref{sect:stars}) are 
          indicated with (s). All velocities are given in the heliocentric frame.}
\label{fig:errHb_err3}
\end{figure}

Velocity measurement from magnesium lines was possible for only 39 objects. In nearly
all of them the best visible was MgI 5183.6 \AA. Unfortunately, for some objects even that 
line could not be well fitted in all spectra, so that their $\bar v_{\rm Mg}$
and $\sigma_{\rm Mg}$ had to be calculated from as few as four spectra. As a result, the 
number of chi-square degrees of freedom was not well determined for the $\sigma_{\rm Mg}$ observable. 
Because of that, the histogram of $\sigma_{\rm Mg}$ in Fig. \ref{fig:sigHb_sigNa} is not directly 
comparable 
to the remaining two histograms. It was plotted for illustrative purpose only, and the 
rms deviations $\sigma_{\rm Mg}$ were not used for further analysis. We only used the 
mean velocities $\bar v_{\rm Mg}$ to check if they differ from their $\bar v_{\rm H}$ 
counterparts. No systematic differences were found, and the mean difference was equal 
to just 5.3~km~s$^{-1}$, which confirmed the reliability of mean-velocity measurements 
based on H${_\beta}$. 

The FXCOR measurements fulfilled the reliability criterion $f_c^m>0.2$ in the case 
of 28 objects, while the stellar Na D$_1$ was strong enough for successful fitting in only 
11 objects. As a result, there were just 9 objects for which we collected four 
complete independent velocity measurements (from H$_\beta$, Mg and stellar Na fitting, 
and from FXCOR). For each spectrum of those objects we calculated the average 
\begin{equation}
 v_4\equiv 0.25(v_{\rm H} + v_{\rm Mg} + v_{\rm Na,s} + v_{\rm FXCOR})
 \label{eq:v4}
\end {equation}
and the corresponding rms deviation $\sigma_4$. Fig. \ref{fig:errHb_err3} shows that 
there are no significant differences between $v_4$ and velocities obtained from 
H$_\beta$ fitting, thus proving the reliability of the latter. All deviations $\sigma_4$ 
except one range between 2 and 20 km s$^{-1}$.

We note that FXCOR returns formal velocity errors, but they are correct only to within 
a scaling factor which depends on the number of counts in the spectra and the Fourier filter 
parameters used \citep[see e.q.][]{for10}. For most of our objects they turned out to
be unrealistically small ($\sim$2 -- 5 km~s$^{-1}$).

\subsection{Velocity variables}
\label{sect:stars}

Fig. \ref{fig:chi_sigHb} suggests that the lower limit of 
$\sigma_H$ for the detection of radial-velocity (rv) variables is $\sim$20~km~s$^{-1}$.
With this in mind, we checked if the velocities of 11 photometric variables
with $\sigma_{\rm H} > 20$ km s$^{-1}$ are compatible with their lightcurves. To that
end we fitted the velocity data of each variable with a sinusoid using its photometric period 
$P_{phot}$ taken from \citet{kal04}. The only exception was V379: for this object a period of 
0.5$P_{phot}$ was used, which is justified given the ambiguity of the lightcurve of \citet{kal04}.

\begin{figure}[t!]
\centering
\includegraphics[width=0.49\textwidth,bb= 50 262 324 689,clip]{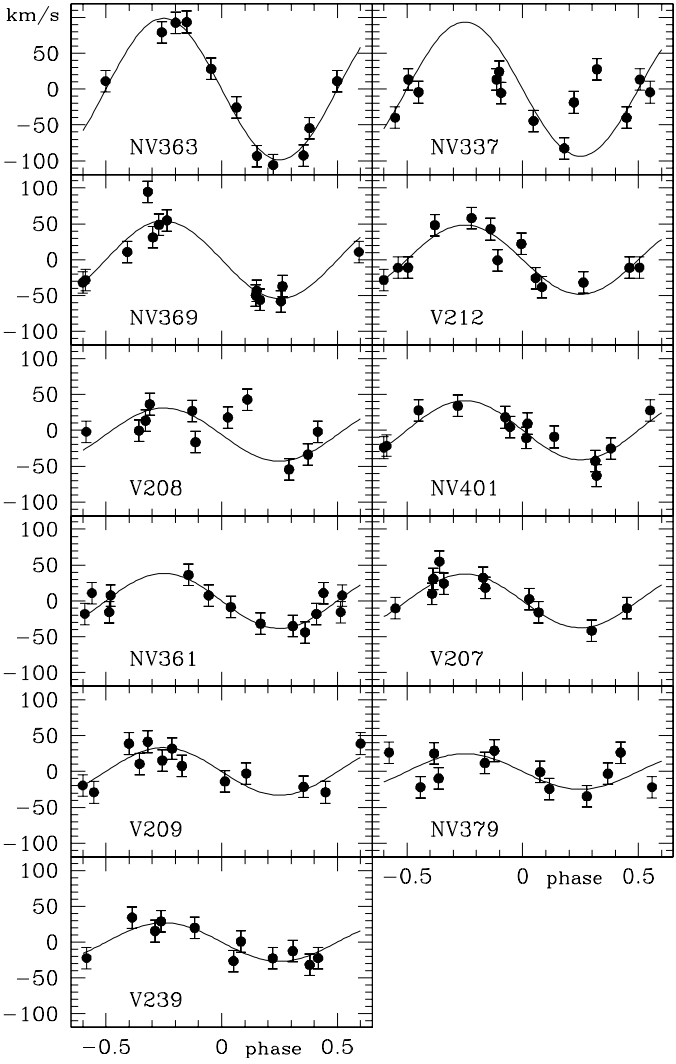}
\caption {Radial velocity curves of photometric variables. For each variable the 
          observational data (points) are phased with the corresponding 
          photometric period $P_{phot}$ taken from \citet{kal04}. The only exception 
          is NV379, where a period of 0.5$P_{phot}$ was used for phasing. Smooth curves 
          are sinusoidal fits. Zero points on vertical axes correspond to systemic velocities. 
          The standard error of velocity measurement, estimated in  
          Sect. \ref{sect:errors}, is 15 km s$^{-1}$.}
\label{fig:pvel_var}
\end{figure}

The fitting was nearly successful - for the total of 110 points only three departed from the fits 
by more than $3\Delta v_{\rm H}$ (two for NV337 and one for V208; see Fig. \ref{fig:pvel_var}). 
In particular, we recovered velocity variations of our test-bed object - the V209 
system, for which we had independent velocity data provided by \citet{kal07a}. From the fit in 
Fig.~\ref{fig:pvel_var} we find that the velocity semiamplitude of V209
is 32.5 km~s$^{-1}$, which well matches the value of 30.7 km~s$^{-1}$ given by 
\citet{kal07a} for the semiamplitude of the primary component (note that 
the primary is four times brighter than the secondary, so that the contribution of the 
secondary to our spectra is rather unimportant). 

Based on Figs. \ref{fig:chi_sigHb}, \ref{fig:sigHb_sigNa} and \ref{fig:pvel_var} 
we conclude that objects with  $\sigma_{\rm H} > 20$ km s$^{-1}$ may be rather safely 
identified as genuine rv-variables. Among straggler candidates witout light-curves
we found 11 such objects
which makes a total of 22 rv-variables in the whole sample. Out of them six satisfied the 
$f^c_m>0.2$ criterion (four photometric variables and two straggler candidates), and in all 
those cases the results from FXCOR confirmed the rv-variability found from line fitting. 
Additionally, we found 17 straggler candidates with $f^c_m>0.2$ showing consistent velocity 
variations whose full amplitude exceeded 30 km~s$^{-1}$, which we consider suspected 
rv-variables (see Table \ref{tab:object_list}). 

\subsection{Radial-velocity membership}
\label{sect:medium}

\begin{figure}[t]
\centering
\includegraphics[width=0.49\textwidth,bb= 28 379 564 690,clip]{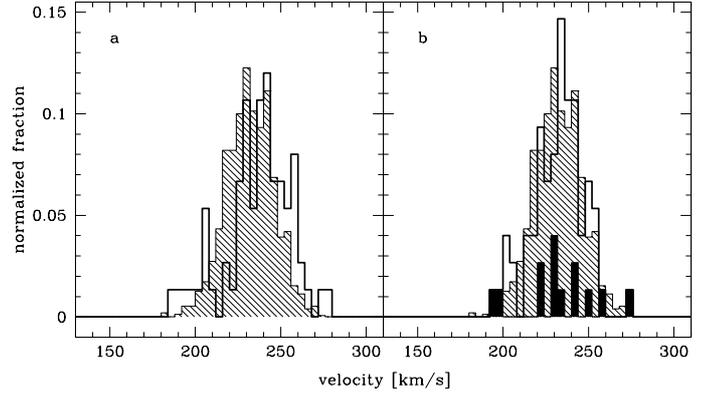}
\caption {Histograms of heliocentric $\bar{v}_{\rm H}$ velocities (heavy line) compared to the 
          histogram of velocities obtained by \citet{vl07} (shaded). a: $\bar{v}_{\rm H}$
          uncorrected; b: $\bar{v}_{\rm H}$ corrected in such a way that the mean velocity 
          of all objects in each VIMOS quadrant is equal to 232.1 km~s$^{-1}$ given recently 
          by \citet{har10} as the systemic velocity of $\omega$ Cen. Black bars in Fig. b 
          indicate red straggler candidates.}
\label{fig:Hb_vel_hist}
\end{figure}

Using the IRAF task RVCORRECT we transferred the velocities $\bar{v}_{\rm H}$ 
to the heliocentric frame. Their histogram is shown in Fig. \ref{fig:Hb_vel_hist}a,
where we compare it with the histogram of velofities measured for over 1500
stars in the field of $\omega$ Cen by \citet{vl07}. Before comparison a constant value
of -6.2 km s$^{-1}$ was subtracted from the latter to account for the difference between 
the systemic velocity found by those authors (238.3 km s$^{-1}$) and the value of 232.1 
km s$^{-1}$ given recently by \citet{har10}.
\footnote{The comparison is allowable because our field of viev and the central 
part of their field of view coincide, and the standard error of $\bar{v}_{\rm H}$, equal 
to $\Delta v_{\rm H}/\sqrt{10}$, is comparable to the velocity error reported in their survey 
($\sim$5 vs. $\sim$8 km~s$^{-1}$).}

\citet{vl07}, who used an instrument with the field of view served by two 
independent CCDs, found that on one of them the mean velocity of all stars $v^\star$ was 
$\sim$9 km~s$^{-1}$ larger than on the other. They corrected for this effect by lowering 
all radial velocities from the first CCD by 4 km s$^{-1}$ and by increasing all radial 
velocities from the second one by an equal amount. Expecting similar problems with our 
four CCDs, we performed the same test. Upon averaging $\bar{v}_{\rm H}$ from Table 
\ref{tab:object_list} over all stars in each quadrant we obtained $v^\star_1=222.5$, 
$v^\star_2=235.5$, $v^\star_3=238.7$ and $v^\star_4=253.1$ km~s$^{-1}$, respectively,
for quadrants 1 -- 4. To account for these differences, we introduced corrections 
\begin{equation}
 \Delta v^\star_i \equiv v_{sys}-v^\star_i,
 \label{eq:vcorr}
\end{equation}
where $v_{sys}$ is the systemic heliocentric velocity of $\omega$ Cen equal to 232.1 km~s$^{-1}$ 
\citep{har10}, and $i$ numbers the quadrants.
We obtained $\Delta v^\star_{1-4}$ = 9.4, -3.4, -6.6 and -21.0 km~s$^{-1}$.

The histogram of corrected velocities is shown in Fig. \ref{fig:Hb_vel_hist}b,
from which it is evident that our correction procedure makes sense.
What remains to be explained are large velocity deviations in quadrants 1 and 4.
Possible explanations include mask flexures (although one would expect their effect to 
be stochastic rather than systematic) and shifts in zero-point velocity caused by 
differences in optical paths of light rays recorded by different CCDs. 

The minimum 
and maximum corrected heliocentric velocity of our targets was equal, respectively, to 
195.1 and 272.4 km~s$^{-1}$ (minimum and maximum uncorrected velocity was, respectively, 
185.4 and 279.6 km~s$^{-1}$).
Based on the kinematic criterion of \citet{vl07}, who assign to $\omega$~Cen all stars with
180 km s $^{-1}\le v \le 300$ km s$^{-1}$, we may say that almost all objects in our 
final sample do indeed belong to the cluster. The membership is firmly excluded  
only for NV369 and B304775, whose velocities, not shown in Fig.~\ref{fig:Hb_vel_hist}, 
fall well below 100 km~s$^{-1}$. Note that this conclusion is valid independently 
of whether we include corrections defined by equation (\ref{eq:vcorr}) or not.

\section{Discussion}
\label{sect:discussion}

The most exciting goal of the present survey -- identification of massive low-luminosity members 
of $\omega$ Cen -- was not reached. The lack of accurate ephemerides prevented us from 
repeating the detailed analysis performed on a smaller sample of $\omega$ Cen photometric 
variables by \citet{roz10}, but no mass function higher than $f_m = 0.22$ was found,
and in addition this particular value was obtained for NV369 -- the only photometric
variable from the sample whose membership was disproved by our analysis. Typical 
$f_m$ for cluster members was below 0.05, indicating rather low-mass companions to 
photometric variables with rv-variations. 

For 72 out of 74 objects investigated here our radial velocity measurements confirmed 
the proper-motion membership of $\omega$ Cen determined by \citet{bel09}. This conclusion 
does not depend on whether corrected or uncorrected velocities are used, as in both cases all
rv-members fulfill the kinematic membership criterion of \citet{vl07}.
The two proper-motion members of $\omega$ Cen which turned out to be field objects are 
B304775 ($mp$ = 100\%) and NV369 ($mp$ = 94\%), whose uncorrected heliocentric velocities
were 44 km~s$^{-1}$ and 81 km s$^{-1}$, respectively. On the other hand, 
the two objects with the lowest proper-motion membership probability, i.e. NV361 ($mp$ = 64\%) 
and NV332 ($mp$ = 84\%), turned out to be genuine members of $\omega$ Cen. As solely in 
those four cases both proper-motion and radial-velocity data were needed to firmly establish 
the status of the object, our results indicate that the proper-motion catalog of \citet{bel09} 
is highly ($\sim$95\%) reliable. 

\begin{figure}
\centering
\includegraphics[width=0.49\textwidth,bb= 58 188 564 689,clip]{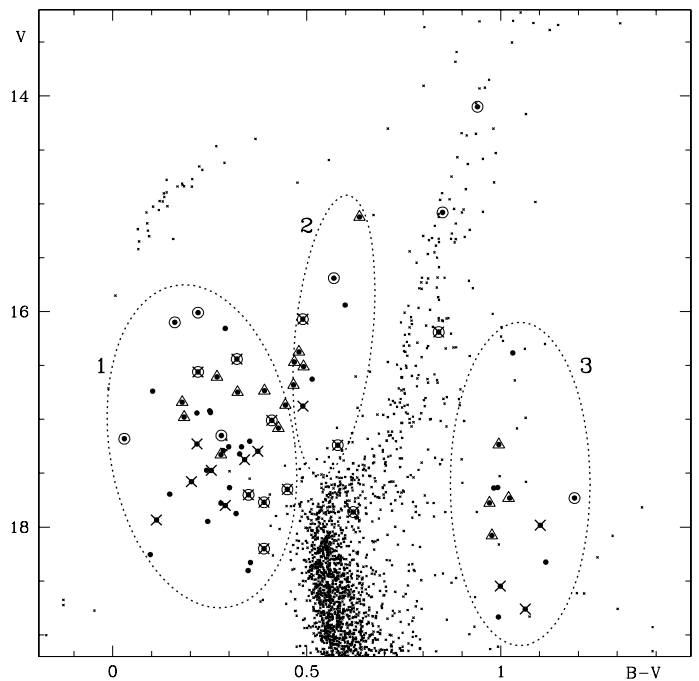}
\caption {Location of the investigated objects (large dots) on the CMD of 
          $\omega$ Cen. Circles: photometric variables. Crosses: radial-velocity variables.
          Triangles: suspected radial-velocity variables. 
          Ellipses 1,~2~and~3 indicate approximate locations of regions occupied, respectively, 
          by blue, yellow and red stragglers (see also Sect. \ref{sect:intro}). The background 
          CM-diagram is from our unpublished photometry of a 2.6$\times$2.6 arcmin field
          centered at $\alpha$ = 13$^h$27$^m$42$^s$.2, $\delta$ = -47$^\circ$23'34'',
          obtained on the 2.5-m duPont telescope at Las Campanas Observatory.} 
\label{fig:cmd_var}
\end{figure}

We found that 11 out of 19 photometric variables and 11 out of 55 straggler candidates
were rv-variables. The possibility that some rv-variable
straggler candidates might be single pulsating stars is rather excluded, as velocity 
variations with $\sigma_{\rm H}>20$ km~s$^{-1}$ would cause detectable photometric 
effects for the \citet{kal04} survey. Thus, our results indicate the binary nature of 22 objects.
In addition, based on cross-correlation
of observed spectra with synthetic templates we identified 17 straggler candidates suspected 
of being rv-variables. All these objects are shown in Fig. \ref{fig:cmd_var} on the 
CMD of $\omega$ Cen.  

The high percentage of binary systems in the area occupied in Fig. \ref{fig:cmd_var} 
by blue and yellow stragglers allows us to firmly state that merging is not the only way 
to produce these objects. In fact, it is quite likely that all stars from our sample located 
in that region are binaries. 
V192 and V205 at ($B-V$, $V$) = (0.22, 16.56), (0.28, 17.15), for 
which we have not found clear rv-variations, are eclipsing variables \citep{kal04}. NV380, 
NV374 and V259 at (0.03, 17.18), (0.16, 16.10) and (0.57, 15.69) have rather long periods 
(7.83, 3.31 and 19.12 d, respectively), and may have easily escaped recognition if their 
orbital inclinations are not close to 90$^\circ$. If we number them among suspected 
rv-variables, for the total of 51 cluster members located in the discussed region there 
will be 33 (i.e. 65\%) for which an indication of rv-variability was found. If we 
additionally include the binary blue stragglers discussed by \citet{roz10}, this ratio 
increases to 69\%. 
 
Similar fractions of binary blue stragglers were found in two old open clusters: 
\citet{matg09} identified 16 binaries among 21 blue stragglers of NGC 188, and \citet{lat07} 
found 8 binaries among 13 blue stragglers of M67. However, both in NGC 188 and M67 almost 
all binary blue stragglers have
much longer periods than the objects discussed here -- typically on the order of $\sim$1000 
days. Our spectroscopic survey lasted two weeks only, and in addition it was not sensitive 
to low velocities. As a result, we would not find any long-period variables even if they
were there, and the lack of such objects in our sample may be a pure selection effect. 
However, the small percentage of short periods in NGC 188 and M67 compared to $\omega$ Cen 
is real, and it must have a physical cause. \citet{matg09} point out that ``the observation 
of binaries with circular orbits and periods of $\sim$1000 d in both NGC188 and M67 indicates 
that these blue straggler binaries have not been disturbed dynamically since their 
formation''. It is thus conceivable that the frequency of dynamical interactions in 
both open clusters is too low to cause efficient tightening of binary orbits, while in 
$\omega$ Cen it might be high enough to produce the observed short-period systems. In any 
case, the fact that binaries are standard members of the blue and yellow straggler population 
is a strong argument in favor of the hypothesis explaining the origin of these objects by a 
significant but rather quiet mass exchange in binary systems which rarely, if ever, leads 
to merging. 
Our findings support the results of \citet{fer106}, who, based on the flat radial 
distribution of blue stragglers in $\omega$ Cen, ruled out the collisional origin of these 
objects, and suggested that they are the progeny of primordial binaries. The same
conclusion was reached by \citet{dal08} in relation to NGC 2419.

Interesting objects have been found also to the right of the 
main sequence and below the subgiant branch of several clusters. They are the 
so-called red stragglers or sub-subgiants \citep[e.g.][]{pla11}. As remarked by \citet{pla11}
in relation to NGC6971, ``these apparent binary stars occupy an area of the CMD ... which is 
not easy to populate with any combination of two normal cluster stars.'' At least some 
candidate red stragglers in $\omega$ Cen are X-ray sources, however the cause of their X-ray 
activity remains obscure \citep{hag10}. The two binary red stragglers in M67 studied 
by \citet{mat03} may belong to the RS CVn class, but their parameters have not been uniquely 
determined. \citet{mat03} and \citet{mat08} argue that their evolutionary history might 
have included mass transfer episodes, mergers, dynamical stellar exchanges and/or close 
encounters leading to envelope stripping. In their conclusion \citet{mat03} point to an
intriguing possibility that these systems are progenitors of blue stragglers. 

According to the kinematic criterion of \citet{vl07}, who define $\omega$~Cen members as   
those with $180 < v < 300$ km s$^{-1}$, all 
the 13 objects from our sample which reside to the right of the main sequence and red giant 
branch of the cluster, are members of 
the cluster (to be on the safe side, one may classify the three objects with extreme 
velocities as ``likely members''; see Fig. \ref{fig:Hb_vel_hist} 
and Table \ref{tab:object_list}). 12 red stragglers are sub-subgiants,
but the last one, B167555 at ($B-V$, $V$) = (1.03, 16.38), is less luminous by only 
$\sim$ 0.2 mag than the photometrically and radial-velocity variable red giant NV379 at 
(0.84, 16.19), so that the alternative term ``red straggler'' fits it much better. 
Our red-straggler sample includes up to eight (i.e. up to 62\%) binaries: the eclipsing 
variable NV332 accompanied by rv-variables B253281, B121450 and B146967 and four suspected 
rv-variables. It is thus conceivable that the remaining 5 objects are also binary systems, which 
would support the final conjecture of \citet{mat03}. If that conjecture is right then red 
stragglers are relatively fresh products of complex interactions between cluster members,
and as such they may turn out to be even more interesting than the blue stragglers themselves.

\section{Conclusions}

Our spectroscopic study of 74 proper-motion members of $\omega$ Cen confirmed the membership 
of 72 objects. Among 55 blue stragglers belonging to our sample we found 22 binaries and 
13 suspected binaries, whereas among 13 red stragglers we found 4 binaries and 4 suspected 
binaries. The fact that binarity is normal in these intriguing populations leads one to expect 
that their 
evolutionary conundrums will be solved once good light-curve and velocity-curve solutions
are found. We hope that our results will prompt the relevant photometric and spectroscopic 
research.

\begin{acknowledgements}
Research of JK, PP, MR and WP is supported by the grant MISTRZ from the Foundation for 
the Polish Science and by the grants N~N203 379936 and N~N203 301335 from the Polish 
Ministry of Science and Higher Education. Support for MC and CC is provided by the Ministry
for the Development, Econmy and Turism Programa Inicativa Cient\'{i}fica Milenio through 
grant P07-021-F, awarded to The Milky Way Millenium Nucleus; by Proyecto Basal PFB-06/2007;
by FONDAP Centro de Astrof\'{i}sica 15010003; and by Proyecto FONDECYT Regular \#1110326.
We are sincerely grateful to the anonymous referee whose detailed suggestions greatly improved 
the quality and the exposition of the paper.
\end{acknowledgements}

\appendix

\section {The baffling histogram of ISM velocity}
\label{sect:disc_medium}

$\omega$ Cen is an ideal laboratory to study small-scale structure and dynamics of the 
interstellar medium (ISM) by using its stars as densely spaced beacons. In fact, within 
the last two decades there were several such attempts.
\citet{ba92} and \citet{wo93,wo94} found several components of NaI D$_2$ at velocities 
between 0 and $-40$ km~s$^{-1}$, ``with the most conspicuous changes of the line profiles
occurring in the most negative components, where variations were seen on an angular scale 
of $\sim$1 arcmin''. Strong patchiness of the absorbing medium was 
reported by \citet{cal05}, who registered ``clumpy extinction variations by a factor of almost 
two across the core of the cluster''. \citet{vl07} detected a {\it redshifted} component of 
the CaII K line. Since it was visible only near the centre of $\omega$ Cen, they suggested 
that it originated in a medium located within the gravitational influence of the cluster.
High positive velocities were also measured in HI and CO emission lines \citep{ori97}.
A detailed mapping of interstellar clouds in front of $\omega$ Cen was recently performed 
by \citet{vl09}. They found the ISM to be patchy at all scales from 0.7$^\circ$ (the size 
of their field) down to about 30\arcsec (the closest distances between sightlines to sampled 
stars). The mean heliocentric velocity of the CaII K and NaI D$_2$ lines obtained from their 
survey is equal to $-24$ and $-11$ km~s$^{-1}$, respectively. The histogram of 
heliocentric NaI D$_2$ 
velocities based on their online data peaks at $-11$ km~s$^{-1}$, and it is practically 
limited to the range between $-6$ and $-16$ km~s$^{-1}$. 

\begin{figure}[h!]
\centering
\includegraphics[width=0.49\textwidth,bb= 35 394 565 692,clip]{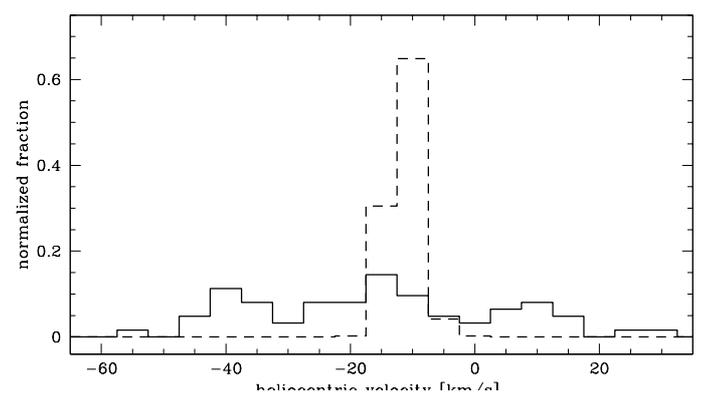}
\caption {Histogram of heliocentric velocities obtained from NaI D$_2$ line. Solid: our 
          data. Broken: \citet{vl09}.}
\label{fig:hist_vel_Na}
\end{figure}

The above brief review indicates that the problem of the ISM in front of $\omega$ Cen is far 
from being solved, and this is what prompted us to extract the relevant information 
from our data. We measured the velocities of the NaI D$_2$ line, converted them to the 
heliocentric reference frame, and added quadrant-dependent corrections introcuced in Sect. \ref{sect:medium}. 
The resulting velocity histogram, shown in Fig. \ref{fig:hist_vel_Na}, turns out to be entirely 
different from that of \citet{vl09}: the central maximum 
is much lower, broader and shifted by $\sim$5 km~s$^{-1}$ toward more negative velocities. 
Moreover, an excess of objects with velocities between $-35$ and $-45$ km~s$^{-1}$ and another 
one centered around +10 km~s$^{-1}$ seem to be present.

\begin{figure}[h!]
\centering
\includegraphics[width=0.49\textwidth,bb= 35 403 562 690,clip]{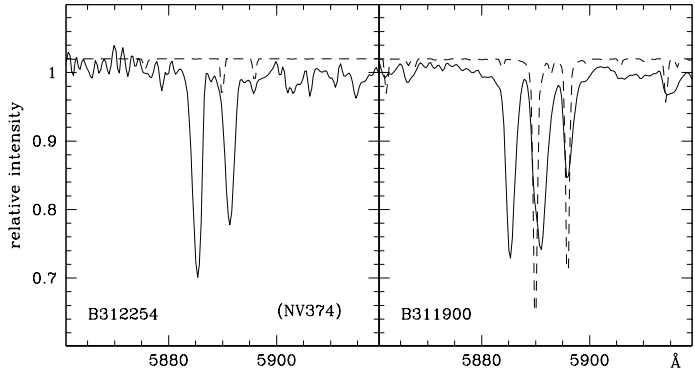}
\caption {NaI doublet in Doppler-corrected spectra of B312254, aka NV374 ($B-V$ = 0.16), and
          B311900 ($B-V$ = 0.64). The observed spectra (solid lines) are compared to synthetic 
          templates assigned as explained in Sect. \ref{sect:stars} (broken lines). The templates 
          are shifted upwards by 0.02 for clarity. Because the stellar spectra have 
          been Doppler-corrected for the velocity obtained from H$_\beta$, the ISM components 
          appear shifted with respect to the rest wavelengths.}
\label{fig:Na_lines}
\end{figure}

Unfortunately, we do not have any common 
objects with \citet{vl09}, so that a direct comparison of the spectra is impossible. Errors 
on our side could originate from wrong velocity measurements, 
misidentification of spectral lines or bad wavelength calibration. However, for each object 
$\bar{v}_{\rm Na}$ is an average over ten spectra. If true velocities were indeed strongly 
peaked at $-11$~km~s$^{-1}$ then it would be difficult to imagine how such mean values could 
be wrong by as much as $\sim$30 km~s$^{-1}$ to produce spurious peaks at $-40$ and $+10$ km~s$^{-1}$. 
Further, the identification of the NaI 
doublet is so easy that we can entirely exclude it as a source of errors. 
In hotter stars there are no 
strong lines with which it could be mistaken, and in cooler stars the doublet turns into 
a unique triplet composed of interstellar D$_2$ and stellar D$_1$ with a blend of 
interstellar D$_1$ and stellar D$_2$ in between (Fig. \ref{fig:Na_lines}). 

Shortwards of H$_\beta$ our wavelentgh calibration is inaccurate in some objects 
\citep{eso11}, but between H$_\beta$ and H$_\alpha$ it should be rather correct. We 
verifed this by selecting a few spectra that contain both these lines, Doppler-correcting 
them for the velocity obtained from H$_\beta$, and comparing with the corresponding synthetic 
templates assigned as explained in Sect. 
\ref{sect:stars}. In Doppler-corrected spectra H$_\alpha$ fell in the right place, 
and the stellar D$_1$ line was also properly placed (Fig. \ref{fig:Na_lines}). Note that 
for a spurious shift of $\sim$30 km~s$^{-1}$ a relatively large calibration error of $\sim$0.3~\AA\ 
is needed which 
should be easily visible in Doppler-corrected spectra. Of course, smaller wavelength 
mismatches could have escaped out attention.

Another factor suspected of generating errors is instrumental effects (mask flexures, different 
zero-point velocities in different quadrants and/or slits). As a result, it 
might turn out that our velocity corrections introduced in Sect. \ref{sect:medium} should vary 
not only between quadrants, but also between slits, and there would be no way to calculate them. 
One would have to accept that absolute velocities obtained from blue VIMOS spectra may be wrong 
by up to $\pm$30 km~s$^{-1}$ (on the other hand, it was shown in Sect. \ref{sect:stars} that the 
instrument is stable enough for reliable measurements of velocity variations with an accuracy 
of $\sim$15 km s$^{-1}$).

 
\begin{longtable}{rrcccrrrrrrcrc}
   \caption{List of objects. \label{tab:object_list}}\\
    \hline
&Bellini &  $B-V$ & $V$  &VQ&$\bar{v}_{\rm H\beta}$&$\sigma_{\rm H\beta}$&
$\bar{v}_{\rm Na}$&$\sigma_{\rm Na}$&Kaluzny&Period&Pvar&Vvar&remark\\
    \hline\hline
    \endfirsthead
    \caption{continued.}\\
    \hline
&Bellini &  $B-V$ & $V$  &VQ&$\bar{v}_{\rm H}$&$\sigma_{\rm H}$&
$\bar{v}_{\rm Na}$&$\sigma_{\rm Na}$&Kaluzny&Period&Pvar&Vvar&remark\\
    \hline\hline
    \endhead
    \hline
    \endfoot
1	&	69263	&	0.36	&	18.33	&	1	&	227.5	&	12.7	&	-5.7	&	9.2	&		&		&		&		&		\\
2	&	69830	&	0.30	&	17.26	&	4	&	239.7	&	12.8	&	25.4	&	13.5	&		&		&		&		&		\\
3	&	81157	&	0.22	&	17.23	&	1	&	210.6	&	34.9	&	-55.7	&	12.8	&		&		&		&	v	&		\\
4	&	85380	&	0.25	&	16.94	&	1	&	230.4	&	12.4	&	-50.8	&	17.1	&		&		&		&		&		\\
5	&	87731	&	0.48	&	16.37	&	4	&	254.5	&	14.0	&	26.0	&	14.9	&		&		&		&	s	&		\\
6	&	95154	&	0.32	&	16.75	&	4	&	256.9	&	15.9	&	2.7	&	6.4	&		&		&		&	s	&		\\
7	&	95702	&	0.11	&	17.93	&	1	&	191.2	&	27.6	&	-49.1	&	20.7	&		&		&		&	v	&		\\
8	&	99931	&	0.29	&	16.16	&	4	&	254.5	&	7.2	&	34.4	&	16.4	&		&		&		&		&		\\
9	&	100977	&	0.39	&	16.73	&	1	&	239.7	&	12.0	&	-43.4	&	13.7	&		&		&		&	s	&		\\
10	&	107036	&	0.27	&	16.61	&	4	&	252.0	&	15.9	&	11.1	&	7.5	&		&		&		&	s	&		\\
11	&	107122	&	0.32	&	17.87	&	1	&	202.9	&	16.6	&	-47.0	&	15	&		&		&		&		&		\\
12	&	111100	&	0.98	&	17.64	&	1	&	185.4	&	11.3	&	-53.2	&	7.8	&		&		&		&		&	lrs	\\
13	&	113098	&	0.35	&	17.20	&	4	&	248.9	&	13.4	&	13.5	&	9.9	&		&		&		&		&		\\
14	&	114490	&	0.28	&	17.78	&	1	&	195.6	&	14.7	&	-48.8	&	10	&		&		&		&		&		\\
15	&	118163	&	0.45	&	17.65	&	1	&	205.5	&	32.0	&	-32.3	&	11.6	&	NV337	&	0.269	&	EW?	&	v	&		\\
16	&	121450	&	1.06	&	18.76	&	4	&	279.6	&	31.8	&	0.0	&	0	&		&		&		&	v	&	rs	\\
17	&	121591	&	0.24	&	17.47	&	1	&	226.5	&	10.5	&	-46.6	&	16.3	&		&		&		&		&		\\
18	&	124330	&	0.49	&	16.07	&	1	&	81.2	&	66.1	&	0.0	&	0	&	NV369	&	1.788	&	?	&	v	&	nm	\\
19	&	130105	&	0.35	&	17.70	&	4	&	243.2	&	30.6	&	9.7	&	13.9	&	NV401	&	0.354	&	?	&	v	&		\\
20	&	131682	&	0.47	&	16.68	&	1	&	197.9	&	10.2	&	-52.1	&	10.7	&		&		&		&	s	&		\\
21	&	133059	&	0.60	&	15.94	&	4	&	266.9	&	13.0	&	16.0	&	14.3	&		&		&		&		&		\\
22	&	135321	&	0.99	&	18.83	&	1	&	232.6	&	10.4	&	-20.6	&	27.3	&		&		&		&		&	rs	\\
23	&	137605	&	1.19	&	17.73	&	4	&	241.4	&	17.0	&	12.7	&	11.3	&	NV332	&	0.247	&	EW	&		&	rs	\\
24	&	140758	&	0.33	&	17.32	&	1	&	242.6	&	12.4	&	-30.9	&	10.6	&		&		&		&		&		\\
25	&	142842	&	0.33	&	17.26	&	4	&	241.4	&	30.7	&	2.4	&	11.2	&		&		&		&		&		\\
26	&	143829	&	0.43	&	17.08	&	1	&	244.3	&	5.3	&	-27.1	&	25.2	&		&		&		&	s	&		\\
27	&	146095	&	0.20	&	17.58	&	4	&	256.9	&	64.1	&	-7.6	&	11.9	&		&		&		&	v	&		\\
28	&	146967	&	1.10	&	17.98	&	1	&	262.7	&	35.6	&	13.1	&	45	&		&		&		&	v	&	lrs	\\
29	&	148727	&	0.03	&	17.18	&	4	&	246.3	&	17.4	&	-4.1	&	15.8	&	NV380	&	7.832	&	?	&		&		\\
30	&	149714	&	0.22	&	16.94	&	1	&	230.5	&	14.6	&	-49.7	&	12	&		&		&		&		&		\\
31	&	154483	&	0.10	&	16.74	&	1	&	228.9	&	13.7	&	-49.4	&	9.9	&		&		&		&		&		\\
32	&	155556	&	0.85	&	15.08	&	4	&	249.6	&	13.7	&	1.2	&	9.7	&	V216	&	23.737	&	LT	&		&	rg	\\
33	&	157688	&	0.25	&	17.95	&	1	&	225.6	&	13.7	&	-40.4	&	10.2	&		&		&		&		&		\\
34	&	159485	&	0.35	&	18.40	&	4	&	263.9	&	11.6	&	0.0	&	0	&		&		&		&		&		\\
35	&	162119	&	0.99	&	17.63	&	1	&	219.7	&	12.6	&	-52.8	&	9.1	&		&		&		&		&	rs	\\
36	&	166068	&	1.00	&	17.23	&	1	&	233.8	&	15.2	&	-49.2	&	10.7	&		&		&		&	s	&	rs	\\
37	&	167555	&	1.03	&	16.38	&	4	&	272.3	&	15.5	&	15.3	&	13.9	&		&		&		&		&	rs	\\
38	&	171825	&	0.51	&	16.63	&	4	&	253.2	&	14.7	&	6.4	&	15.1	&		&		&		&		&		\\
39	&	176840	&	0.25	&	16.92	&	1	&	229.2	&	15.8	&	-43.7	&	8.4	&		&		&		&		&		\\
40	&	177256	&	0.57	&	15.69	&	4	&	237.3	&	17.6	&	-2.4	&	9	&	V259	&	19.120	&	sp?	&		&		\\
41	&	179912	&	0.32	&	16.44	&	1	&	229.8	&	23.0	&	0.0	&	0	&	V239	&	1.189	&	EA	&	v	&		\\
42	&	180526	&	0.97	&	17.78	&	4	&	250.4	&	16.2	&	-35.0	&	11.6	&		&		&		&	s	&	rs	\\
43	&	234296	&	0.15	&	17.69	&	3	&	251.2	&	7.8	&	0.0	&	0	&		&		&		&		&		\\
44	&	234998	&	0.39	&	18.20	&	2	&	237.4	&	24.5	&	-2.1	&	15.5	&	NV361	&	0.682	&	EA	&	v	&		\\
45	&	241994	&	0.30	&	17.63	&	3	&	227.8	&	14.0	&	7.3	&	17.5	&		&		&		&		&		\\
46	&	244408	&	0.41	&	17.01	&	2	&	248.0	&	28.9	&	-12.3	&	14.8	&	V208	&	0.306	&	EW	&	v	&		\\
47	&	249955	&	0.49	&	16.51	&	2	&	222.0	&	13.4	&	-13.9	&	15.7	&		&		&		&	s	&		\\
48	&	250383	&	0.34	&	17.37	&	3	&	255.4	&	21.5	&	0.0	&	0	&		&		&		&	v	&		\\
49	&	253226	&	1.02	&	17.73	&	3	&	237.8	&	9.3	&	0.0	&	0	&		&		&		&	s	&	rs	\\
50	&	253281	&	1.00	&	18.55	&	2	&	225.2	&	22.1	&	-24.0	&	25.4	&		&		&		&	v	&	rs	\\
51	&	258274	&	0.49	&	16.88	&	2	&	206.9	&	27.0	&	-21.3	&	17.3	&		&		&		&	v	&		\\
52	&	258539	&	0.18	&	16.84	&	3	&	241.0	&	10.8	&	19.3	&	8.4	&		&		&		&	s	&		\\
53	&	262766	&	0.94	&	14.10	&	2	&	246.1	&	18.4	&	-11.3	&	22	&	NV390	&	15.710	&	?	&	s	&	rg	\\
54	&	263560	&	0.37	&	17.30	&	3	&	207.3	&	18.4	&	9.8	&	14.3	&		&		&		&	v	&		\\
55	&	265591	&	0.29	&	17.29	&	2	&	219.1	&	19.7	&	0.0	&	0	&		&		&		&		&		\\
56	&	269775	&	0.45	&	16.87	&	3	&	235.7	&	11.5	&	17.4	&	8	&		&		&		&	s	&		\\
57	&	274338	&	0.18	&	16.98	&	2	&	237.3	&	6.8	&	-13.1	&	17	&		&		&		&	s	&		\\
58	&	277621	&	0.98	&	18.08	&	3	&	241.9	&	16.7	&	-27.6	&	10.2	&		&		&		&	s	&	rs	\\
59	&	279427	&	0.29	&	17.80	&	2	&	239.7	&	30.5	&	31.7	&	16.7	&		&		&		&	v	&		\\
60	&	281267	&	0.10	&	18.26	&	3	&	259.5	&	17.5	&	0.0	&	0	&		&		&		&		&		\\
61	&	287148	&	0.28	&	17.33	&	3	&	230.0	&	15.0	&	16.0	&	11.2	&		&		&		&	s	&		\\
62	&	289620	&	0.25	&	17.47	&	2	&	256.7	&	34.1	&	0.0	&	0	&		&		&		&	v	&		\\
63	&	293439	&	0.58	&	17.24	&	2	&	243.9	&	31.9	&	-7.7	&	16.9	&	V212	&	2.467	&	EA	&	v	&		\\
64	&	297708	&	0.39	&	17.77	&	2	&	238.5	&	26.6	&	-18.1	&	14.5	&	V207	&	0.276	&	EW	&	v	&		\\
65	&	297731	&	0.22	&	16.56	&	3	&	256.5	&	22.6	&	15.7	&	11.3	&	V209	&	0.834	&	EA	&	v	&		\\
66	&	302520	&	0.28	&	17.15	&	3	&	230.9	&	8.6	&	0.0	&	0	&	V205	&	0.369	&	EA	&		&		\\
67	&	304775	&	0.52	&	15.77	&	3	&	44.4	&	15.9	&	35.0	&	8.9	&		&		&		&		&	nm	\\
68	&	306843	&	1.12	&	18.33	&	3	&	206.0	&	14.8	&	23.0	&	6.5	&		&		&		&		&	lrs	\\
69	&	311900	&	0.64	&	15.12	&	2	&	228.5	&	17.0	&	-11.9	&	11.1	&		&		&		&	s	&		\\
70	&	312254	&	0.16	&	16.10	&	3	&	246.7	&	15.1	&	18.3	&	6.7	&	NV374	&	3.311	&	?	&		&		\\
71	&	317269	&	0.47	&	16.47	&	2	&	240.6	&	14.5	&	-11.6	&	14	&		&		&		&	s	&		\\
72	&	319396	&	0.22	&	16.01	&	2	&	241.9	&	17.0	&	-12.7	&	11.1	&	V192	&	1.376	&	EA	&		&		\\
73	&	321474	&	0.62	&	17.86	&	3	&	256.6	&	74.6	&	7.5	&	16.4	&	NV363	&	0.824	&	EA	&	v	&	to	\\
74	&	328605	&	0.84	&	16.19	&	3	&	233.8	&	19.8	&	14.9	&	6.9	&	NV379	&	7.104	&	?	&	v	&	rg	\\

\end{longtable}
    \noindent
    Legend: \\
    Bellini - object's number in \citet{bel09}. \\
    $B-V$ - color index, not corrected for reddening.\\
    $V$ - $V$-band magnitude.\\
    VQ - number of VIMOS quadrant. \\
    $\bar{v}_{\rm H}$ and $\sigma_{\rm H}$ - 
    mean value and rms deviation of velocity calculated from ${\rm H\beta}$. \\
    $\bar{v}_{\rm Na}$ and $\sigma_{\rm Na}$ - mean value and rms deviation of 
    velocity calculated from the interstellar NaI D$_2$ line (0 if the spectrum did not reach 
     to the NaI doublet).\\
    Kaluzny - object's number in \citet{kal04}.\\
    Period - photometric period in days. \\
    Pvar - type of photometric variability after \citet{kal04}:
    EA = Algol; EW = W UMa; LT = long term; SP = spotted.\\
    Vvar - flag for velocity variability: v = variable; s = suspected. \\
    Remarks: rs = red straggler; lrs = likely red straggler; rg = red giant; to = turnoff object; 
    blank - object located to the left of main sequence and red giant branch (blue straggler, 
    yellow straggler or hot subdwarf).\\
    The velocities are given in the heliocentric frame without corrections described in 
    Sect. \ref{sect:medium}. The values of the corrections are 9.4, -3.4, -6.6 and 21.0 km~s$^{-1}$
    for quadrants 1 -- 4, respectively.

\end{document}